\documentstyle[prl,aps,epsf]{revtex}
\draft
\begin{document}
\twocolumn[\hsize\textwidth\columnwidth\hsize\csname @twocolumnfalse\endcsname

\title{Superconducting Vortex Logic Antidots}
\author{ C. J. Olson Reichhardt$^{1,4}$, 
C. Reichhardt$^{2,4}$, and B. Jank{\' o}$^{3,4}$} 
\address{$^1$T-12 and $^2$Center for Nonlinear Studies, 
Los Alamos National Laboratory,  Los Alamos, New Mexico 87545}
\address{
$^3$Department of Physics, University of Notre Dame, Notre Dame, 
Indiana 46617 }
\address{ 
$^4$Materials Science Division, Argonne National Laboratory, Argonne, Illinois
60429}
\date{\today}
\maketitle

\begin{abstract}
We examine a building block for logic devices in which
the positions of superconducting vortices in coupled elongated antidots
provide the elementary logic states of $0$ and $1$. 
We show analytically and through simulation the maximum operating
frequency of a pair of antidots as a function of antidot spacing and
elongation.  At finite temperatures, a signal can propagate through
a series of identically shaped antidots with correctly chosen spacing,
with an exponential distribution of switching times for the signal
to move over by one antidot.
\end{abstract}

\vskip2pc]
%\vskip2pc
\narrowtext
Recently there has been considerable interest in superconductors with 
artificial pinning arrays, such as artificial arrays 
of holes in 
thin-film type-II superconductors \cite{Fiory,Baert,Reichhardt}. 
Under a magnetic field,
the flux penetrates the film in the form of individual quantized vortices which
become pinned by the holes. Experiments \cite{Fiory,Baert} 
and simulations \cite{Reichhardt} have shown that 
for fields where the number of vortices equals an integer multiple  of 
the number of holes, peaks or anomalies in the transport and critical current
occur. These anomalies are correlated with the formation of highly
ordered vortex lattice crystals.
Effective periodic pinning arrays
can also be constructed by placing magnetic dots on superconductors,
which produces strong pinning and 
commensuration effects \cite{Martin}, or by
fabricating large ``blind'' holes (or antidots)
which do not pass completely through
the film but instead modulate the film thickness.
In this case multiple 
vortices can be trapped at an individual antidot
\cite{Bezryadin} and the vortex positions from one antidot are correlated
with the positions in neighboring antidots \cite{Multi}.
Recently experiments
have also demonstrated that nanodots of superconducting materials can 
capture individual vortices \cite{Geim,Peeters}.    
Besides circular
pinning sites, it is also possible to 
fabricate
elongated pinning sites \cite{Metlushko}. 
In this case an individual vortex may not necessarily be
located at the center of the pinning site but 
will reside along a line that cuts through
the center of the site in the long direction.  

Since the vortices have specific arrangements at certain integer matching or
fractional matching fields, it should be possible to use 
the locations of the vortices or the flux configurations as 
elementary logic states such as 0 and 1.
These states can then
be measured or propagated by the application of a magnetic field
or current. 
Puig {\it et al.} \cite{Puig} 
made one initial proposal along these lines, in which
a superconducting island with $2\times2$ plaquettes 
was considered as a logic element.
In this case the minima in the resistance can be associated with different
flux configurations.   

The quantum-dot cellular automata (QCA) is
another system that uses the locations or
configurations of the particles as basic logic states
\cite{QCA,QCA2}. 
A basic cell consists of four quantum dots containing
two localized electrons.
Due to the Coulomb repulsion between the electrons,
there are two possible ground state configurations with the
electrons located at the diagonals of the cell,
slanted toward the right or left. These two states provide the fundamental
logic units. With different geometrical arrangements of the basic cell, 
various logic devices can be constructed.   
In the QCA system the logic states are propagated by quantum-mechanical means.
So far, elementary QCA systems have been demonstrated to operate only at 
extremely low temperatures. Since these systems must be adiabatically 
switched between logic states, 
signal propagation times are relatively slow.    

In this paper we examine the building blocks for a  
superconducting vortex logic system \cite{Hastings} 
proposed in analogy with the 
quantum cellular automata.
In our model we consider a superconductor that has been nanofabricated
to contain elongated pinning sites, where each site captures 
a single vortex.
Our basic unit consists of coupled elongated antidots 
of identical shape where, due to the
repulsion of the vortices, two ground states can be formed. 
We show that by flipping one vortex in a pair of antidots,
the vortex in the other antidot also flips, and we
examine the response time of the second vortex as a function of the
antidot geometry.  We show that at finite temperatures, 
a signal can be propagated through a series of 
identically shaped antidots that have
been fabricated in a specific geometry.

\begin{figure}
\center{
\epsfxsize=3.5in
\epsfbox{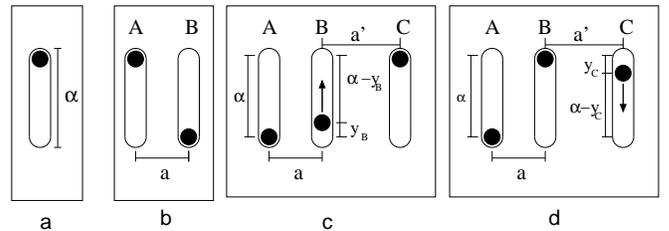}}
\caption{Geometry of the vortex logic elements.  Open shapes represent the
antidots, and filled circles represent the vortices.  (a) A single antidot
of elongation $\alpha$.  The vortex is shown in logic state 1, at the
top of the antidot.  Logic state 0 is represented when the vortex is
at the bottom of the antidot.
(b) Vortices A and B in neighboring antidots
separated by distance $a$ assume opposite logic states.
(c) An example of a signal propagating through three antidots.  Vortex A
has been switched to logic state 0. Vortex B 
is in the process of switching, and has moved a distance $y_B$.  
The spacing between 
antidots A and B is $a$, and the spacing between antidots B and C is
$a^\prime > a$.
(d) Vortex C is in the process of switching, and has moved a distance
$y_C$.}
\label{fig:123dots}
\end{figure}

The individual logic element in our system is a single vortex inside
an elongated antidot, 
a pinning site
created through nanolithography techniques. 
As illustrated in Fig.~\ref{fig:123dots}(a),  
we take the
elongation to be in the $y$ direction.  In the presence of a
neighboring antidot, parallel to
the first antidot and offset in the $x$ direction,
the two equilibrium positions of the vortex are at
either end of the antidot.  Thus, to create our logic element, we define
a vortex position at the top of the antidot to be a logic '1', 
as shown in Fig.~\ref{fig:123dots}(a), while
a vortex positioned at the bottom of the antidot is in a logic '0' state.
One way in which the logic state of the antidot can be externally controlled
is by means of an STM tip made of magnetic material.  If this tip is 
moved in the $y$ direction at a given frequency, it should be 
possible to drag the vortex in the antidot below between the logic
'0' and '1' states.  The position of a vortex in a given antidot can
also be detected by using an STM in spectroscopy mode.  The vortex
location can be identified due to the fact that the spectrum at the
vortex core is very different from both superconducting or normal
spectra \cite{Hess}. 

In order to transmit information through the system,   
neighboring antidots must be coupled.
We first consider an isolated pair of antidots
a distance $a$ apart, illustrated in Fig.~\ref{fig:123dots}(b).  
The vortices are prepared in an equilibrium 
configuration where vortex A
is in logic state 1, and vortex B is in logic state 0.  At time $t=t_0$
the position of vortex A is flipped by external means to be in logic
state 0.  For operation of the vortex logic, we require that 
vortex B will subsequently flip without external intervention
to logic state 1, so that the final state of the system is reversed
from the initial state.

In the case of two isolated antidots, vortex B does not need to 
overcome an energy barrier in order to flip.  We can 
estimate the operating frequency of the two-vortex system by
finding the time required to complete the flip.
The interaction between the two vortices 
in a thin film is given by the Pearl interaction \cite{Pearl},
which can be written 
\begin{equation}
f^{vv}(r)=\frac{\Phi_0^2}{2\mu_0 \pi \lambda^2}\frac{d}{r}
\end{equation}
in the limit $r \ll 2\lambda^2/d$.  Here, $\Phi_0$ is the elementary
flux quantum, $\mu_0$ is the permeability of free space, $\lambda$ is
the London penetration depth of the superconductor, and $d$ is the
thickness of the superconducting film.  
The vortices obey overdamped dynamics given by
${\bf f}_i=\eta{\bf v}_i$, where $\eta=B_{c2}\Phi_0/\rho_{N}$, $B_{c2}$ is
the upper critical field, and $\rho_{N}$ is the normal state
resistivity.
We assume that the antidots are spaced $a$ apart in the $x$ direction,
and are of length $\alpha$ in the elongated $y$ direction.  
If the position of vortex A is switched and held fixed such that both
vortices are in the same logic state, vortex B
requires a transit time $t_{tr}$ before it reaches the opposite
side of the antidot and the system returns to equilibrium.  The
transit time can be written
\begin{equation}
t_{tr} = \int_{\delta}^{\alpha} \frac{1}{v(y)}dy
\end{equation}
where $v(y)={\bf f}\cdot{\hat y}/\eta$, 
the $y$ velocity of vortex B at position $y$.
The integration must start from a small offset $\delta$ because
if the two vortices are at the same $y$ location, they exert
no force on each other in the $y$ direction and will not move
without thermal assistance.  This means that in an experiment
conducted at low temperatures where thermal fluctuations are
insignificant, vortex A must be moved past the $y$ position of
vortex B by a distance $\delta$.
Putting in the approximation for
the thin film interaction gives
\begin{equation}
\frac{t_{tr}}{t_0}=\frac{1}{2}(\alpha^2-\delta^2)+a^2\ln\left(\frac{\alpha}{\delta}\right)
\end{equation}
where time is measured in units of 
$t_0 = \eta/f_0^\prime$,
with $f_0^\prime = \Phi_0^2/(2\pi\mu_0\lambda^2)$.
Distances are measured in units of the film thickness $d$.
Thus decreasing the spacing $a$ or the antidot length $\alpha$ will
produce faster switching.

We compare this theoretical result to the transit times of the
vortex obtained from a two-dimensional numerical simulation
with open boundary conditions.
The equation of motion for a vortex $i$ is 
\begin{equation}
{\bf f}_{i} = {\eta} {\bf v}_{i}  = {\bf f}^{vv}_{i}
+ {\bf f}^{p}_{i} + {\bf f}^{T}_i  
\end{equation}
The Langevin force from the temperature is ${\bf f}^{T}_i$ and has the
properties $<f^{T}(t)> = 0$ and 
$<f^{T}_i(t)f^{T}_j(t^{'})> = 2\eta k_{B}T\delta_{ij}\delta(t - t^{'})$.  
Initially we consider the case $T=0$.
The force ${\bf f}_{i}^{p}$  
is from the pinning well, 
represented by an ordinary
parabolic trap that has been split in half and elongated in the
$y$ direction.  There is no $y$ direction confining force in
the central elongated portion of the pin.  The maximum pinning
force is $f_{p}$ in both the $x$ and $y$ directions.  The 
radius of the pin is $r_{p}$, and the total length of the pin in
the $y$ direction is $\alpha+2r_p$.  
We first consider two wells a distance $a$ apart, each containing
a single vortex. 

\begin{figure}
\center{
\epsfxsize=3.5in
\epsfbox{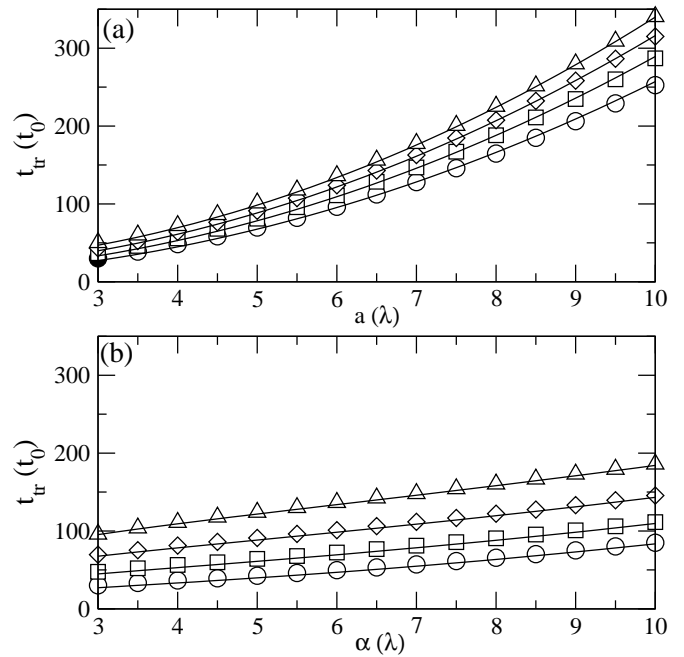}}
\caption{
Transit time $t_{tr}$ for vortex B to cross the antidot and
move from one logic state to the other after vortex A has been switched.
(a) $t_{tr}$ as a function of antidot spacing $a$ for fixed
$\alpha=$ 3 (circles), 4 (squares), 5 (diamonds), and 6 (triangles), 
for a two-well system with $T=0$, $r_p=0.24\lambda$, 
$f_p=0.4f_0^\prime$, and $\delta=0.24\lambda$.
The symbols represent transit times measured in simulations, while
the lines are plots of Eq. 3.
(b) $t_{tr}$ from simulation and Eq. 3 for the same system as
a function of antidot anisotropy $\alpha$ with fixed
$a=$  3 (circles), 4 (squares), 5 (diamonds), and 6 (triangles). 
}
\label{fig:transit}
\end{figure}

Figure~\ref{fig:transit}(a) 
shows the transit time $t_{tr}/t_0$ obtained from simulation
(symbols)
as a function of $a$ with $T=0$, $r_p=0.24\lambda$, $f_p=0.4f_0^\prime$, 
$\delta=0.24\lambda$,
and $\alpha/\lambda= 3$, 4, 5, and 6,
along with corresponding plots of Eq.~3 (solid lines).  
In Fig.~\ref{fig:transit}(b) we show $t_{tr}/t_0$ as a function of $\alpha$ for
$a/\lambda=3$, 4, 5, and 6.
In each case we find excellent agreement
between simulation and theory.
In the case of a Nb film of thickness 2000 \AA \, 
with antidots of anisotropy $\alpha=3\lambda=$135 nm, and
with spacing between the dots $a=3\lambda=$135 nm,
the transit time of $t_{tr}=27.2t_0$ from Eq.~3
with $\delta=0.24\lambda=$10.8 nm
corresponds to an actual time of 1.4 ns, 
indicating that the
maximum operating frequency for Nb antidots of this size and geometry
is 696 MHz.  Smaller or more closely spaced dots will operate at
higher frequencies.

To create any type of device, the logic states need to be propagated
over distances further than a single well. We therefore consider
the case of three antidots, A, B, and C, illustrated in 
Fig.~\ref{fig:123dots}(c).  
All of the antidots are the same
shape and size, but the antidot spacing varies, so that antidots A and B are
separated by $a$, but antidots B and C are separated by $a^{\prime}$.
If $a^{\prime}=a$, then when the vortex in dot A is switched externally,
vortex B will not be able to switch because it experiences a potential
barrier due to the presence of vortex C.  Instead, the new minimum
energy for vortex B will be at the center of well B.  This is
undesirable for logic operations.  In order to allow vortex B to
switch, we must increase $a^{\prime}>a$ to reduce the energy barrier
produced by the repulsion
from vortex C, and we must also
introduce thermal fluctuations to allow thermal activation over the
energy barrier.  In a low-temperature material such as Nb, 
the thermal fluctuations may
remain prohibitively small below $T_c$ unless very small antidots are
fabricated.  In a high-temperature superconductor
such as BSCCO, it is much easier to produce thermal activation
even for relatively large antidots.

For the configuration shown in Fig.~\ref{fig:123dots}(c), 
where vortex A has been switched
to a new logic state but vortex B has not yet switched,
we can write an expression for the $y$ position
of vortex B at which the net $y$ force on vortex B from vortices A and C
is zero.  We rescale all distances by $\alpha$, so that
$\tilde y = y/\alpha$, $\tilde a = a/\alpha$, and 
$\tilde a^{\prime} = a^{\prime}/\alpha$, and we take $\tilde y = 0$ as 
the starting position of vortex B and $\tilde y=1$ as the final
switched position.  We obtain
\begin{equation}
2 {\tilde y}^3_B - 3 {\tilde y}^2_B + 
(1 + {\tilde a}^{\prime 2} + {\tilde a}^2)
{\tilde y}_B - {\tilde a}^2 = 0.
\end{equation}
After vortex B reaches $\tilde y_B$ satisfying this expression, it has
crossed the potential barrier and can move freely to the opposite side of
the well.  Similarly, after vortex B has switched to the new
logic state [Fig.~\ref{fig:123dots}(d)], we can write
an expression for the $y$ position at which the force on vortex C from 
vortices A and B in the switched state is zero:
\begin{equation}
2 {\tilde y}^3_C - 3 {\tilde y}^2_C + 
(1 + {\tilde a}^{\prime 2} + 2 {\tilde a}{\tilde a}^{\prime}
+ 2{\tilde a}^2) {\tilde y}_C - {\tilde a}^{\prime 2} = 0.
\end{equation}
The position $\tilde y_C$ marks the end of the potential energy
barrier.  

Vortices B and C can cross their respective energy barriers 
by thermal activation.
Well C should be placed as close as
possible to well B in order to enhance the coupling of vortex C to
vortex B, so $a^{\prime}$ should be made as small as possible.  However, 
as $a^{\prime}$ approaches $a$, the coupling between vortices
A and B is weakened, and vortex B will never switch initially.
Thus $a^{\prime}$ should be chosen just slightly larger than the value
at which vortex B can no longer switch.  
For $\tilde a = 0.187$, 
vortex B ceases to switch below $\tilde a^\prime_{crit}=0.29$;
thus, for optimal
operation, $\tilde a^\prime$ should be chosen slightly
above $\tilde a^\prime_{crit}$.

The thermal fluctuations must be large enough for vortices B and
C to reach positions $\tilde y_B$ and $\tilde y_C$ at
$\tilde a^\prime_{crit}$ by thermal activation in order for
the signal to propagate.  The vortices also thermally fluctuate
in the $x$ direction, and therefore
if the antidots are wide in the $x$ direction, 
the value of $a^{\prime}_{crit}$ is increased compared to the value
obtained from Eqn. 5 since the minimum possible $x$ distance  
$a^\prime_{\rm eff}$ between vortices B and C is less than 
$a^\prime$.  The following relation holds: 
$a^\prime-2r_p < a^\prime_{\rm eff} < a^\prime$.  The same is true of
$a$.  Thus in experiments the pins must be fabricated slightly
further than $a$ or $a^\prime$ apart.
Additionally, since the switching of the
vortices is now thermally activated, strict clocking of the signal
is {\it no longer possible} as it was in the case of two wells.
A mechanism to obtain strict clocking has been demonstrated
in Ref.~\cite{Hastings}.

\begin{figure}
\center{
\epsfxsize=3.5in
\epsfbox{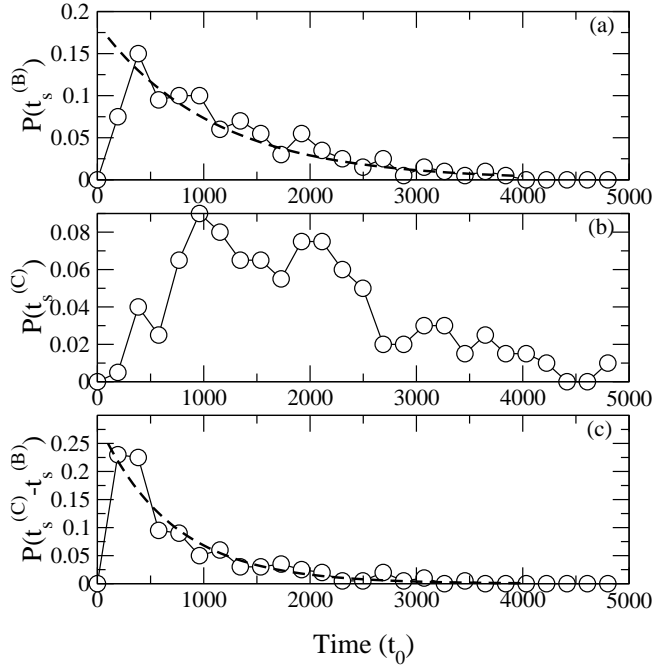}}
\caption{
(a) The distribution of switching times $P(t_s^{(B)})$ 
for vortex B obtained from
200 independent simulations of a system with
$f_T=1.2f_0^{\prime}$, $a=0.5\lambda$, $a^\prime=0.68\lambda$,
and $\alpha=2.48$.  Time is measured from $t=0$.
The dashed line indicates an exponential distribution with parameter
$1/\lambda_e=1078 t_0$, the mean value ${\bar t}_s^{(B)}$.
(b) $P(t_s^{(C)})$ for vortex C, with time measured from $t=0$.
The mean switching time is ${\bar t}_s^{(C)}=1771 t_0.$
(c) $P(t_s^{(C)}-t_s^{(B)})$ 
for vortex C, with time measured from $t=t_s^{(B)}$ for
each run, the time at which vortex B switched.
The dashed line indicates an exponential distribution with
parameter $1/\lambda_e=693 t_0$, which is ${\bar t}_s^{(C)}-{\bar t}_s^{(B)}$.
}
\label{fig:dist}
\end{figure}

The distance between the wells required for the operation 
of the three well system scales with $\alpha$.  The
smaller $\alpha$ is, the smaller the distance between pins can be
made and still allow operation of all three wells.  
We consider the case of BSCCO where thermal activation can play a
significant role below $T_c$.  
In Fig.~\ref{fig:dist} 
we show the distribution of switching times $P(t_s)$ obtained from
200 runs with different random temperature seeds.  
Here $f_T=1.2f_0^{\prime}$, $a=0.5\lambda$, $a^\prime=0.68\lambda$,
and $\alpha=2.48$. 
Fig.~\ref{fig:dist}(a) shows that
$P(t_s^{(B)})$ for vortex B is exponentially distributed
with mean value ${\bar t}_s^{(B)}=1078 t_0$.  
$P(t_s^{(C)})$ for vortex C, with
time measured from $t=0$ before vortex B has switched, is plotted in
Fig.~\ref{fig:dist}(b), 
and the distribution is clearly broader and more heavily
weighted toward later times.  $P(t_s^{(C)})$ for vortex C is merely the
product of two exponential distributions, as can be seen from 
Fig.~\ref{fig:dist}(c), where
we plot the switching time of vortex C with time measured from 
$t=t_s^{(B)}$, the switching time of vortex B.  
$P(t_s^{(C)}-t_s^{(B)})$ is also exponentially distributed
with a mean value of 693$t_0$.

More than three wells can be connected, but the spacing between
well $n$ and $n+1$ must always be greater than the spacing between wells
$n$ and $n-1$.  Thus there is a practical limitation on the total length
of device that can be fabricated in this fashion; 
when the wells are spaced too far
apart, the vortices will thermally decouple.  
Additionally, as illustrated in Fig.~\ref{fig:dist}, 
the distribution of switching
times for the final well will become increasingly broad 
and approach a Gaussian as the number
of wells is increased.  Regardless of the number of wells, these
logic elements operate only from the narrowly spaced end to the 
widely spaced end.  A
signal introduced at the widely spaced end will not be able to
propagate to the narrow end.  If all the wells are spaced equally,
it is still possible for excitations to propagate through the wells
under thermal activation if the temperature is high enough.  However,
these excitations move diffusively in either direction and are not
well controlled.  It is possible instead to employ a ratchet mechanism
to obtain controlled, clocked motion of the signal \cite{Hastings}.

In summary we have examined the basic building blocks for
a vortex cellular automata 
that is analogous to a quantum-dot cellular automata. 
We consider 
elongated pinning sites in superconducting samples where there is 
one vortex per antidot. 
The vortices can form two
ground states with the vortices located at diagonals in order to minimize
their interaction energy.  We obtained analytically and in simulation
the maximum frequency of operation for a two well system.
For the simplest pipeline geometry of three wells
we find that, for finite temperatures, a change in logic state introduced
at the first well can be propagated over specified distances.

{\it Acknowledgments---}
We thank W. Kwok and V.V. Metlushko  for useful discussions.
This work was supported by the U.S. DoE, Office of Science, under
Contract No. W-31-109-ENG-38.
CJOR and CR were supported by the U.S. DoE under Contract No.
W-7405-ENG-36. BJ was supported by NSF-NIRT award DMR02-10519.
and the Alfred P. Sloan Foundation.

\end{document}